\documentclass[twoside,letterpaper]{soups-poster} 
\usepackage{graphicx}
\usepackage{amsmath}
\begin{document}
%
% --- Author Metadata here ---
%\conferenceinfo{Symposium on Usable Privacy and Security(SOUPS)}{2010, July 14--16, 2009, Redmond, WA, USA}
%\CopyrightYear{2010} % Allows default copyright year (200X) to be over-ridden - IF NEED BE.
%\crdata{0-12345-67-8/90/01}  % Allows default copyright data (0-89791-88-6/97/05) to be over-ridden - IF NEED BE.
% --- End of Author Metadata ---

\title{User-Centred Security Education: A Game Design to Thwart Phishing Attacks}
%
% You need the command \numberofauthors to handle the 'placement
% and alignment' of the authors beneath the title.
%
% For aesthetic reasons, we recommend 'three authors at a time'
% i.e. three 'name/affiliation blocks' be placed beneath the title.
%
% NOTE: You are NOT restricted in how many 'rows' of
% "name/affiliations" may appear. We just ask that you restrict
% the number of 'columns' to three.
%
% Because of the available 'opening page real-estate'
% we ask you to refrain from putting more than six authors
% (two rows with three columns) beneath the article title.
% More than six makes the first-page appear very cluttered indeed.
%
% Use the \alignauthor commands to handle the names
% and affiliations for an 'aesthetic maximum' of six authors.
% Add names, affiliations, addresses for
% the seventh etc. author(s) as the argument for the
% \additionalauthors command.
% These 'additional authors' will be output/set for you
% without further effort on your part as the last section in
% the body of your article BEFORE References or any Appendices.

\numberofauthors{1} %  in this sample file, there are a *total*
% of EIGHT authors. SIX appear on the 'first-page' (for formatting
% reasons) and the remaining two appear in the \additionalauthors section.
%
\author{
% You can go ahead and credit any number of authors here,
% e.g. one 'row of three' or two rows (consisting of one row of three
% and a second row of one, two or three).
%
% The command \alignauthor (no curly braces needed) should
% precede each author name, affiliation/snail-mail address and
% e-mail address. Additionally, tag each line of
% affiliation/address with \affaddr, and tag the
% e-mail address with \email.
%
% 1st. author
\alignauthor
Nalin Asanka Gamagedara Arachchilage\titlenote{}\\
       \affaddr{Australian Centre for Cyber Security}\\
       \affaddr{University of New South wales at the Australian Defence Force Academy}\\
      % \affaddr{Vancouver, BC Canada}\\
       \email{nalin.asanka@adfa.edu.au}
% 2nd. author
\iffalse
\alignauthor
Ivan Flechais\titlenote{}\\
       \affaddr{University of Oxford}\\
       \affaddr{Wolfson Building, Parks Road}\\
       \affaddr{Oxford, UK, OX1 3QD}\\
       \email{ivan.flechais@cs.ox.ac.uk}
% 3rd. author
\alignauthor 
Konstantin Beznosov\titlenote{}\\
       \affaddr{University of British Columbia}\\
       \affaddr{2332 Main Mall}\\
       \affaddr{Vancouver, BC Canada}\\
       \email{beznosov@ece.ubc.ca}
\and  % use '\and' if you need 'another row' of author names
% 4th. author
\fi
}

\maketitle

\section{Introduction}
Security exploits can include cyber threats such as computer programs that can disturb the normal behaviour of computer systems (viruses), unsolicited e-mail (spam), malicious software (malware), monitoring software (spyware), attempting to make computer resources unavailable to their intended users (Distributed Denial-of-Service or DDoS attack), the social engineering, and online identity theft (phishing) \cite{arachchilage2014security}. One such cyber threat, which is particularly dangerous to computer users is phishing \cite{dhamija2006phishing} \cite{arachchilage2011design} \cite{arachchilage2014security}. Phishing is well known as online identity theft, which aims to steal sensitive information such as username, password and online banking details from its victims. Automated anti-phishing tools have been developed and used to alert users of potentially fraudulent emails and websites. However, these tools are not entirely reliable in detecting phishing attacks \cite{sheng2007anti} \cite{arachchilage2013game}. Even the best anti-phishing tools missed over 20 percent of phishing websites \cite{zhang2007cantina}. Because ``human'' is the weakest link in information security \cite{purkait2012phishing} \cite{arachchilage2013game} \cite{arachchilage2014security}. It is not possible to completely avoid the end-user, for example in personal computer use, one mitigating approach for computer and information security is to educate the end-user in security prevention \cite{arachchilage2013game} \cite{truong2006storyboarding} \cite{sheng2007anti} \cite{zhang2007cantina} \cite{downs2007behavioral} \cite{arachchilage2014security}. 

The aim of this research study focuses on a design and development of a game prototype for mobile platforms to educate individuals about phishing attacks. Therefore, the study asks how does one identify which issues the game design needs to be addressed? The elements of a game design framework developed by Arachchilage and Love \cite{arachchilage2013game} for avoiding phishing attacks were used to address the game design issues. Our mobile game design aimed to enhance the users' avoidance behaviour through their motivation to protect themselves against phishing threats. Garera et al. \cite{garera2007framework} strongly argue it is often possible to differentiate phishing websites from legitimate ones by carefully looking at the URL. Therefore, this mobile game prototype designed to teach people to identify legitimate URLs from mimic ones. 

A think-aloud study was conducted, along with a pre- and post-test, to assess the game design framework though the developed mobile game prototype. The study results showed a significant improvement of participants' phishing avoidance behaviour in their post-test assessment. Furthermore, the study findings suggest that participants' threat perception, safeguard effectiveness, self-efficacy, perceived severity and perceived susceptibility elements positively impact threat avoidance behaviour, whereas safeguard cost had a negative impact on it.

\section{Game Design Issues}
To answer these issues, the elements of a game design framework \cite{arachchilage2013game} were incorporated into the mobile game prototype context. The game design framework (Figure 1) describes individual computer users' behaviour in avoiding the threat of malicious information technologies such as phishing attacks \cite{arachchilage2013game}. The framework examined how individuals avoid phishing threats by using given anti-phishing game based education.

\begin{figure} 
\centering
\includegraphics [height=2.5in,width=3.5in]{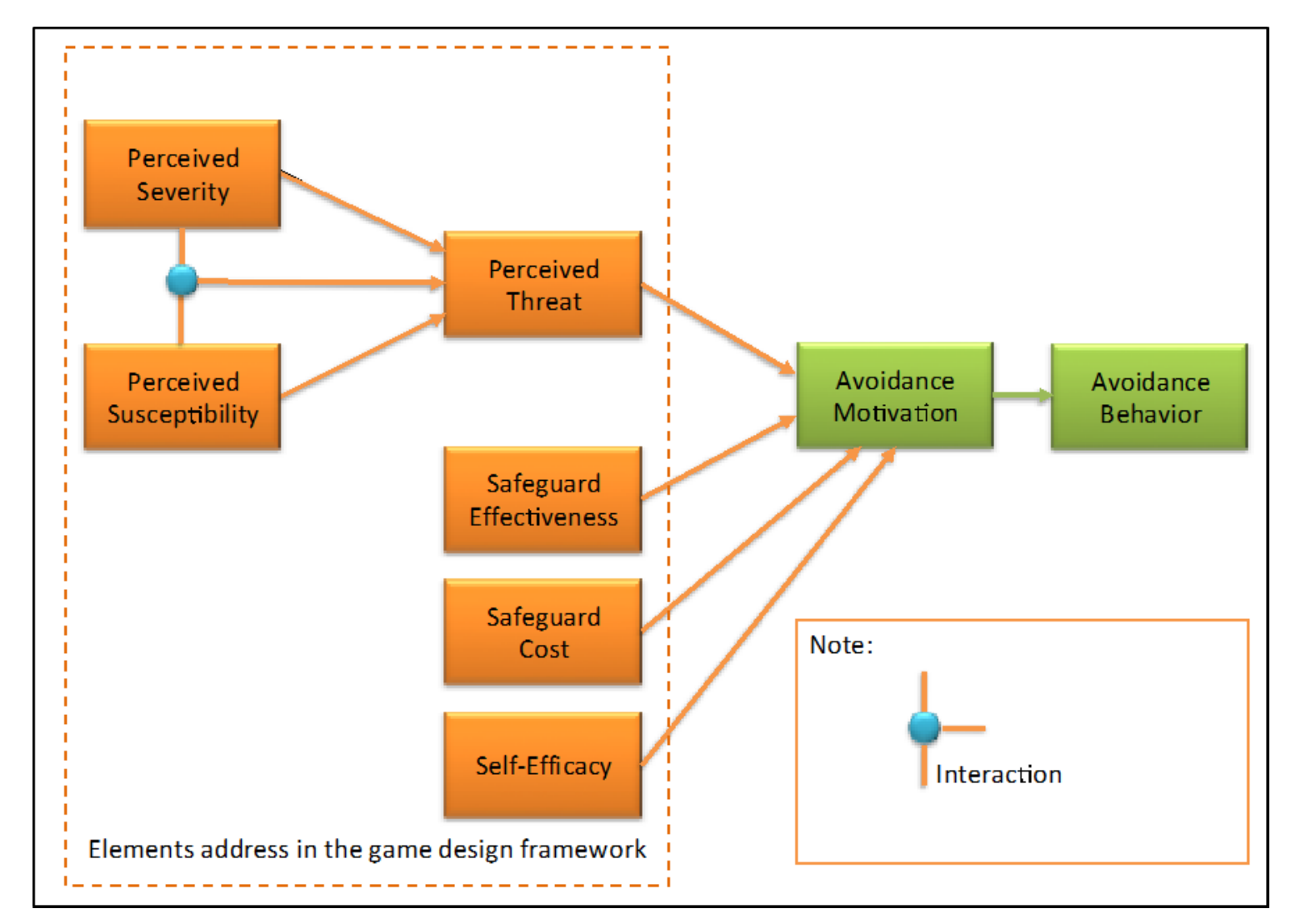}
\caption{The game design framework \protect\cite{arachchilage2013game}}
\vskip -6pt
\end{figure}

Consistent with the game design framework (Figure. 1) \cite{arachchilage2013game}, the users' phishing threat avoidance behaviour is determined by avoidance motivation, which, in turn, is affected by perceived threat. Perceived threat is influenced by perceived severity and susceptibility as well as their combination. Users' avoidance motivation is also determined by the three constructs such as safeguard effectiveness, safeguard cost, and self-efficacy. In addition, the game design framework posits that perceived threat is influenced by the combination of perceived severity and susceptibility. Whilst the game design framework informs the issues that the game design needs to address, it should also indicate how to structure this information and present it in a game context. Therefore, the game design based on a story attempts to develop threat perceptions, making individuals more motivated to avoid phishing attacks and use safeguarding measures. Finally, the elements of the game design framework were incorporated into the mobile game prototype to enhance individuals' phishing threats avoidance behaviour through their motivation to protect themselves from phishing attacks.

\section{Story and Mechanism}
Storytelling techniques are used to grab attention, which can also help to focus on interesting aspects of reality \cite{arachchilage2014storyboardgame}. Stories can be based on personal experiences or famous tales or they could also be aimed at building a storyline that associates content units, inspires, or reinforces.

The game prototype is based on a scenario of a character of a small fish and `his' teacher who live in a big pond. The more appropriate, realistic and content relevant the story, the better the chances that it will trigger users. The main character of the game is the small fish, who wants to eat worms to become a big fish. The game player roll plays as a small fish. However, he should be careful of phishers those who try to trick him with fake worms. This represents phishing attacks by developing threat perception in the game storyboard design. Each worm is associated with a website address (URL), which appears in a dialog box. The game was designed with a total of 10 URLs to randomly display including five good worms and five bad worms. The small fish's job is to eat all the real worms which associate legitimate website addresses and reject fake worms which associate with fake website addresses before the time is up. This attempts to develop the severity and susceptibility of the phishing threat in the game storyboard design. 
The other character is the small fish's teacher, who is a matured and experienced fish in the pond. If the worm associated with the URL is suspicious and if it is difficult to identify, the small fish can go to `his' teacher and request help. The teacher could help him by giving some tips on how to identify bad worms. For example, ``website addresses associate with numbers in the front are generally scams,'' or ``a company name followed by a hyphen in a URL is generally a scam". Whenever the small fish requests help from the teacher, the score will be reduced by certain amount (in this case by 100 seconds) as a payback for safeguard measure. This attempts to address the safeguard effectiveness and the cost needs to pay for the safeguard in the game storyboard design.
 
The game prototype consists of total 10 URLs to randomly display worms including five good worms (associated with legitimate URLs) and five fake worms (associated with phishing URLs). If the user correctly identified all good worms while avoiding all fake worms by looking at URLs, then he will gain 10 points (in this case each attempt possible to score 1 point). If the user falsely identified good worms or fake worms, each attempt loses one life out of total lives remaining to complete the game. If the user requested help from the big fish (in this case small fish's teacher) each attempt loses 100 seconds out of total remaining time to complete the game, which is 600 seconds. Therefore, self-efficacy of preventing from phishing attacks will be addressed in the game storyboard design when the user comes across throughout the game. 

\section{Game Design}
The game was initially sketched in a storyboard using ink pen, post-it notes, and papers based on the above mentioned story \cite{arachchilage2014storyboardgame} \cite{truong2006storyboarding}. To explore the viability of using a game for preventing phishing attacks, a working prototype model was developed for a mobile telephone on Android platform and which is shown in Figure 2. Because most significant feature of a mobile environment is ``mobility" itself such as mobility of the user, mobility of the device and mobility of the service \cite{parsons2006study}. It enables users to be in contact while they are outside the reach of traditional communicational spaces \cite{boyinbode2015mobilect}. For example, one can play a game on his mobile device while travelling on the bus or train, or waiting in a queue.  

\begin{figure}
\centering
\includegraphics [height=2.5in,width=3.5in]{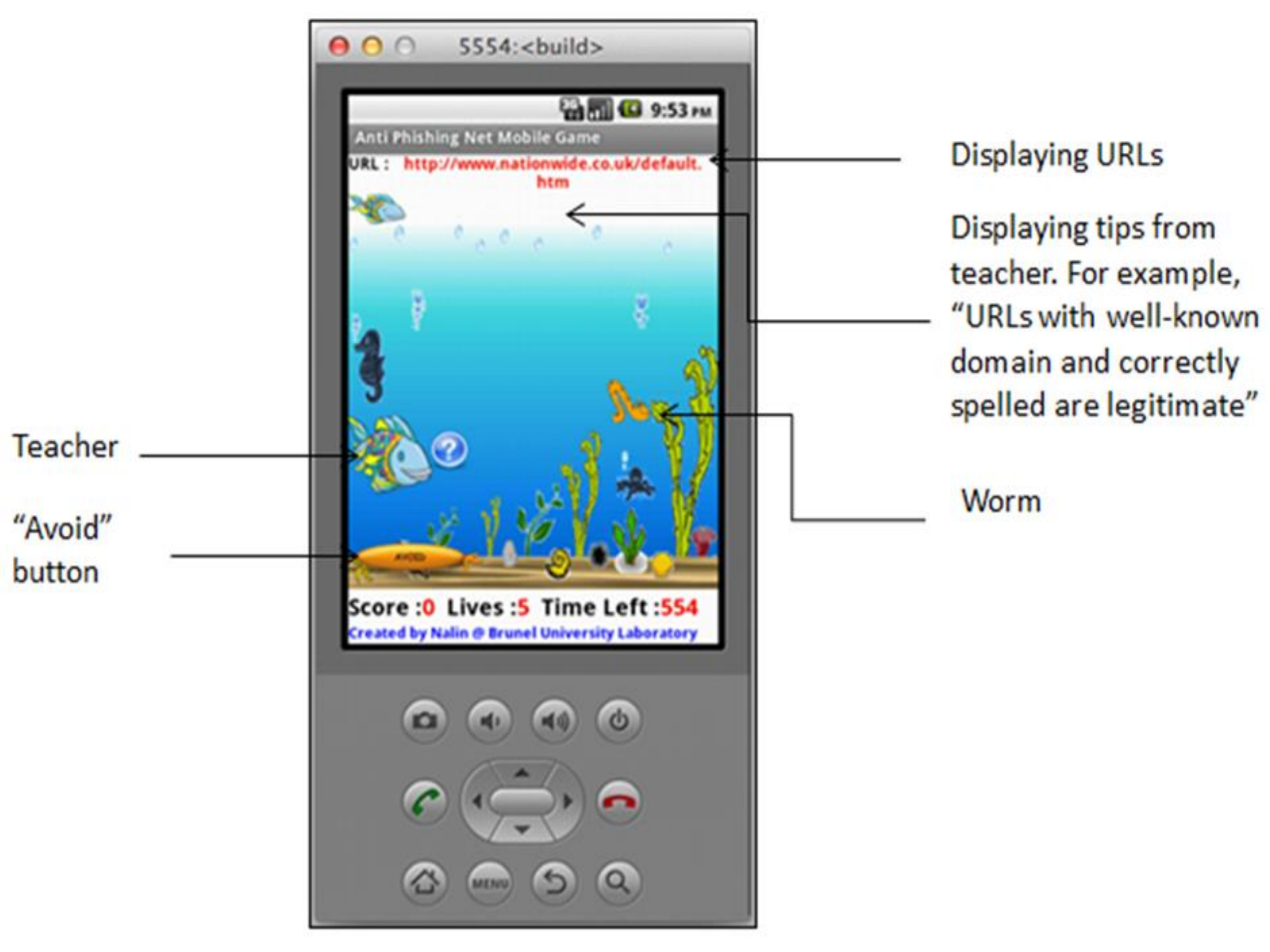}
\caption{The mobile game prototype on Android platform}
\vskip -6pt
\end{figure}

The player is given instructions before starting the game. Then the main menu of the mobile game prototype appears, along with underwater background and the corresponding sound effects. A light water bubbling sound is played in the background throughout the game to make the user feel that they are in the pond. A URL is displayed with each worm; where the worms are randomly generated.

If the worm associated with URL is legitimate, then the user is expected to tap on the worm in order to increase their score. However, if the user fails to identify the legitimate URL, then remaining lives will be reduced by one point. On the other hand, if the worm associated with the URL is phishing, then the user is also expected to tap on the ``AVOID" button to reject the URL, in order to increase the score. If the user fails to do this, then remaining lives will be reduced by one point. If the worm associated with the URL is suspicious and if it is difficult to identify, the user can tap on the big fish (in this case, teacher fish) to request help. Then some relevant tips will be displayed just below the URL. For example, ``website addresses associate that have numbers in the front are generally scams." Whenever the user taps on the big fish, the time left is reduced by 100 points (in this case 100 seconds). Finally, the user gains 10 points if all given URLs were correctly identified within 5 lives and 600 seconds.  

\section{Results}

The current study empirically evaluated the game design framework introduced by Arachchilage and Love (2013) through a prototype of an educational mobile game. A think-aloud study was conducted, along with a pre- and post-test, to assess the game design framework. The study used 20 participants with each one participating for approximately one-hour. 

Initially, we evaluated the participants subjective satisfaction of the mobile game prototype using SUS (System Usability Scale) scoring approach introduced by Brooke \cite{brooke1996sus}. The score was significantly high, 83.62 out of 100 ($\approx 84\%$) \cite{brooke1996sus}. The research study employed Paired-samples t-test to compare the means scores for the participants' pre- and post-tests (Pallant, 2007). The results are encouraging. Participants, who played the mobile game, scored 56\% in the pre-test and 84\% in the post-test. There was a statistically significant increase in the post-test ((Pre-test: M= 56.00, SD=17.911 and Post-test: M=84.00, SD=13.139), t(19)= -7.97, p<0.005 (two-tailed)).

There was a significant improvement of 28\% of the participants' phishing avoidance behaviour in the post-test (p<0.005 (two-tailed)). Eighteen participants scored above 80\%, whilst five of them scored full marks (100\%) in the post-test. All participants scored above 50 percent in their post-test.  The individual participant's score during their engagement with the mobile game prototype is shown in Figure 3. It has been seen that a considerable improvement of overall participants' phishing avoidance behaviour through the mobile game prototype. Finally, the study findings revealed that participants' threat perception, safeguard effectiveness, self-efficacy, perceived severity and perceived susceptibility elements positively impact their threat avoidance behaviour, whereas safeguard cost had a negative impact on it. 

\begin{figure}
\centering
\includegraphics [height=1.5in,width=3.5in]{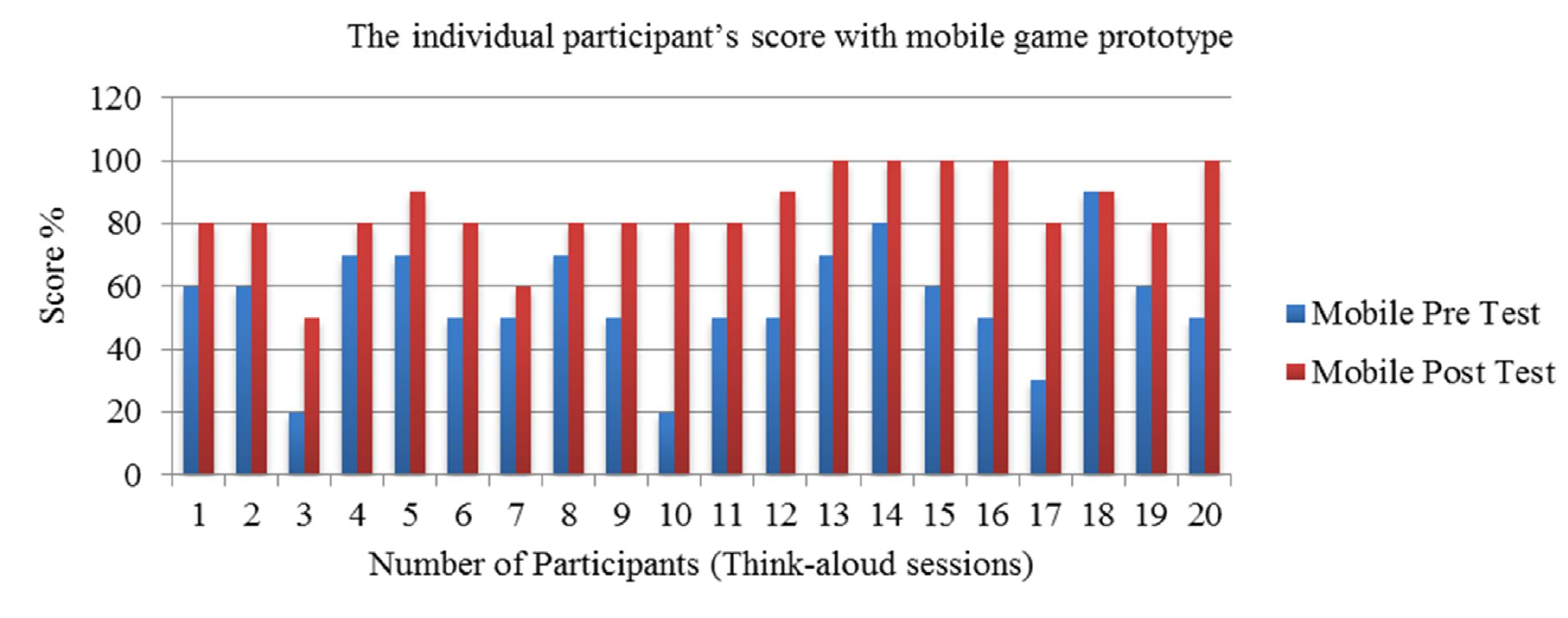}
\caption{The individual participant's score during their engagement with the mobile game prototype.}
\vskip -6pt
\end{figure}
 
\section{Conclusion}
This research focuses on a design and development of a game for mobile platforms to educate computer users to thwart phishing attacks. It asks how does one identify which issues the game design prototype needs to be addressed? The elements of a game design framework developed by Arachchilage and Love \cite{arachchilage2013game} for avoiding phishing attacks were incorporated into the mobile game prototype design context. The objective of our proposed game prototype was to teach user how to identify phishing website addresses (URLs).The study employed SUS, as the first step to assess the subjective satisfaction of mobile game prototype interface. Then, a think-aloud study was conducted along with a pre- and post-test in order to evaluate the game design framework. The current study findings revealed that the overall mobile game prototype enhanced user avoidance behaviour by motivating them to protect themselves from phishing attacks. 

%
% The following two commands are all you need in the
% initial runs of your .tex file to
% produce the bibliography for the citations in your paper.
%{\footnotesize
\bibliographystyle{abbrv}
\bibliography{sigproc}  % sigproc.bib is the name of the Bibliography in this case

\begin{thebibliography}{10}

\bibitem{arachchilage2014storyboardgame}
N.~A. Arachchilage, I.~Flechais, and K.~Beznosov.
\newblock A game storyboard design for avoiding phishing attacks.
\newblock {\em Proceedings of the 11th Symposium On Usable Privacy and Security
  (SOUPS)}, 2014.

\bibitem{arachchilage2011design}
N.~A.~G. Arachchilage and M.~Cole.
\newblock Design a mobile game for home computer users to prevent from phishing
  attacks.
\newblock In {\em Information Society (i-Society), 2011 International
  Conference on}, pages 485--489. IEEE, 2011.

\bibitem{arachchilage2013game}
N.~A.~G. Arachchilage and S.~Love.
\newblock A game design framework for avoiding phishing attacks.
\newblock {\em Computers in Human Behavior}, 29(3):706--714, 2013.

\bibitem{arachchilage2014security}
N.~A.~G. Arachchilage and S.~Love.
\newblock Security awareness of computer users: A phishing threat avoidance
  perspective.
\newblock {\em Computers in Human Behavior}, 38:304--312, 2014.

\bibitem{boyinbode2015mobilect}
O.~Boyinbode and D.~Ng'ambi.
\newblock Mobilect: an interactive mobile lecturing tool for fostering deep
  learning.
\newblock {\em International Journal of Mobile Learning and Organisation},
  9(2):182--200, 2015.

\bibitem{brooke1996sus}
J.~Brooke.
\newblock Sus-a quick and dirty usability scale.
\newblock {\em Usability evaluation in industry}, 189(194):4--7, 1996.

\bibitem{dhamija2006phishing}
R.~Dhamija, J.~D. Tygar, and M.~Hearst.
\newblock Why phishing works.
\newblock In {\em Proceedings of the SIGCHI conference on Human Factors in
  computing systems}, pages 581--590. ACM, 2006.

\bibitem{downs2007behavioral}
J.~S. Downs, M.~Holbrook, and L.~F. Cranor.
\newblock Behavioral response to phishing risk.
\newblock In {\em Proceedings of the anti-phishing working groups 2nd annual
  eCrime researchers summit}, pages 37--44. ACM, 2007.

\bibitem{garera2007framework}
S.~Garera, N.~Provos, M.~Chew, and A.~D. Rubin.
\newblock A framework for detection and measurement of phishing attacks.
\newblock In {\em Proceedings of the 2007 ACM workshop on Recurring malcode},
  pages 1--8. ACM, 2007.

\bibitem{parsons2006study}
D.~Parsons, H.~Ryu, and M.~Cranshaw.
\newblock A study of design requirements for mobile learning environments.
\newblock In {\em null}, pages 96--100. IEEE, 2006.

\bibitem{purkait2012phishing}
S.~Purkait.
\newblock Phishing counter measures and their effectiveness--literature review.
\newblock {\em Information Management \& Computer Security}, 20(5):382--420,
  2012.

\bibitem{sheng2007anti}
S.~Sheng, B.~Magnien, P.~Kumaraguru, A.~Acquisti, L.~F. Cranor, J.~Hong, and
  E.~Nunge.
\newblock Anti-phishing phil: the design and evaluation of a game that teaches
  people not to fall for phish.
\newblock In {\em Proceedings of the 3rd symposium on Usable privacy and
  security}, pages 88--99. ACM, 2007.

\bibitem{truong2006storyboarding}
K.~N. Truong, G.~R. Hayes, and G.~D. Abowd.
\newblock Storyboarding: an empirical determination of best practices and
  effective guidelines.
\newblock In {\em Proceedings of the 6th conference on Designing Interactive
  systems}, pages 12--21. ACM, 2006.

\bibitem{zhang2007cantina}
Y.~Zhang, J.~I. Hong, and L.~F. Cranor.
\newblock Cantina: a content-based approach to detecting phishing web sites.
\newblock In {\em Proceedings of the 16th international conference on World
  Wide Web}, pages 639--648. ACM, 2007.

\end{thebibliography}
%\bibliography{references}
% You must have a proper ".bib" file
%  and remember to run:
% latex bibtex latex latex
% to resolve all references
%
% ACM needs 'a single self-contained file'!
%
%APPENDICES are optional
%\balancecolumns
\end{document}